\documentclass[letterpaper]{article} 
\usepackage{aaai23}  
\usepackage{times}  
\usepackage{helvet}  
\usepackage{courier}  
\usepackage[hyphens]{url}  
\usepackage{graphicx} 
\def\UrlFont{\rm}  
\usepackage{natbib}  
\usepackage{caption} 
\usepackage{algorithm}
\usepackage{algorithmic}

\usepackage{newfloat}
\usepackage{listings}
\DeclareCaptionStyle{ruled}{labelfont=normalfont,labelsep=colon,strut=off} 
\floatstyle{ruled}
\newfloat{listing}{tb}{lst}{}
\floatname{listing}{Listing}
\title{AI Audit: A Card Game to Reflect on Everyday AI Systems }

\author {
    Safinah Ali\textsuperscript{\rm1}, Vishesh Kumar\textsuperscript{\rm2}, Cynthia Breazeal\textsuperscript{\rm1}
}
\affiliations{
    \textsuperscript{\rm1}Massachusetts Institute of Technology, \\ \textsuperscript{\rm2}Northwestern University  \\
    \textsuperscript{\rm1}\{safinah, cynthiab\} @ media.mit.edu, \textsuperscript{\rm2}vishesh.kumar@northwestern.edu
}
\begin{document}

\maketitle

\begin{abstract}
An essential element of K-12 AI literacy is educating learners about the ethical and societal implications of AI systems. Previous work in AI ethics literacy have developed curriculum and classroom activities that engage learners in reflecting on the ethical implications of AI systems and developing responsible AI. There is little work in using game-based learning methods in AI literacy. Games are known to be compelling media to teach children about complex STEM concepts. In this work, we developed a competitive card game for middle and high school students called “AI Audit” where they play as AI start-up founders building novel AI-powered technology. Players can challenge other players with potential harms of their technology or defend their own businesses by features that mitigate these harms. The game mechanics reward systems that are ethically developed or that take steps to mitigate potential harms. In this paper, we present the game design, teacher resources for classroom deployment and early playtesting results. We discuss our reflections about using games as teaching tools for AI literacy in K-12 classrooms.
\end{abstract}

\section{Introduction}
Tools and systems powered by Artificial Intelligence (AI) algorithms have become an undeniable part of our everyday lives. Children frequently interact with AI-powered systems, upload their data on them and are affected by them. This makes it timely and imperative to teach children about the fundamentals of how AI works, where it is used and what the social and ethical implications of AI are. The AI4k12 initiative that was established by the Association for the Advancement of Artificial Intelligence (AAAI) and the Computer Science Teachers Association (CSTA) has determined the “Big 5 ideas in AI” which determine what students should know about AI and be able to do with it~\cite{touretzky2019envisioning}. These ideas include perception - how computers perceive the world using sensors; representation and reasoning - how agents maintain representations of the world and use them for reasoning; learning - how computers learn from data; natural interaction - how intelligent agents require many kinds of knowledge to interact naturally with humans; and social impact - how AI can impact society in both positive and negative ways. Societal impact is an especially essential big idea in AI since it not only protects children from the negative societal impacts of AI by raising awareness, but also educates the future makers of AI systems to design responsible AI, maximizing positive societal impacts and think critically about the potential harms. Several AI systems that have significant negative societal impact are already used by middle and high schoolers. For instance, social media apps used by children can use targeted advertising algorithms to manipulate their buying behavior. This makes learning about social and ethical implications of these systems during these years even more critical. 

There is a lot of previous work focusing on ethical AI education for middle and high schoolers. Researchers have developed curricula primarily on ethics in AI~\cite{ali2019constructionism} or used an integrated ethics approach where the societal implications of algorithms were discussed while students learned about technical constitution or applications of the AI systems~\cite{zhang2022integrating, williams2022ai+}. Among pedagogical methods, previous work has used teaching lectures where instructors discuss how certain AI systems have impacted society, or unplugged activities where students reflect on who stakeholders of the technology are and how they could be positive or negatively impacted, or project-based learning approaches where students designing an algorithm reflect on its societal implications and minimize harm.

In this work, we introduce a new game-based learning approach of teaching children the societal impact of AI systems using a card game. Games are known to be an effective learning method that can facilitate discussion in an engaging manner. We designed a card game called “AI Audit” that encourages children to critique everyday AI systems with potential harms while simultaneously thinking of ways to mitigate the said harm. This is a competitive card game where children play as business owners of various AI powers tools and systems. Players can set up new businesses from their “Business” cards, and other players can challenge their businesses with associated “Harm” cards. Challenges can be countered by “Feature” cards that are mitigation steps players can take to design their systems more responsibly and taking necessary steps to counter the potential societal implications of the system. The goal of the game is to spark reflective conversation on everyday AI systems and raise awareness about their societal implications that students may not be aware. 

In this paper, we describe the design of the game, how it can be played and initial playtesting results. We also reflect on the use of game-based AI ethics learning approaches and using games to facilitate the discussion of AI systems as socio-technical systems. This game is an AI learning resource contribution that can be used in addition to existing AI curricula or as a stand-alone game for children. 

\section{Background}
\subsection{K-12 AI Learning}
Given the prevalence of AI in children’s life and tools that they interact with, and the relevance of AI in future careers, the last decade has seen a surge in AI literacy courses, tools and teaching programs for K-12 students. In 2018, AAAI and CSTA developed national guidelines for teaching AI to K-12 students, where they outlined 5 big ideas that all students must know:  computers perceive the world using sensors, agents maintain models/representations of the world and use them for reasoning, computers can learn from data, making agents interact comfortably with humans is a substantial challenge for AI developers, and AI applications can impact society in both positive and negative ways~\cite{touretzky2019envisioning}. 

There are several existing curriculum efforts for engaging k12 students in AI learning, including, Code.org’s AI for Oceans program that focus on Machine Learning (ML) concepts and the ethical use of AI, Zhorai, a conversational agent for children to explore ML concepts\cite{lin2020zhorai}, Scientist-inSchools program’s curriculum from Australia covering basic AI concepts, AI vocabulary, and the history of AI~\cite{heinze2010action}, IRobot’s AI curriculum for high school students~\cite{burgsteiner2016irobot}, International Society for Technology in Education (ISTE) AI curriculum for high school students and MIT AI Education Initiative’s collection of AI curricula and tools which include the “Developing AI Literacy (DAILy)” curriculum for middle school children~\cite{lee2021developing}, and the “AI+Ethics” curriculum~\cite{ali2019constructionism}, the “How to Train your Robot” curriculum that uses project-based learning of ML concepts~\cite{williams2021train}. To aid young AI learners with limited mathematics or programming expertise, educators and researchers have developed interactive learning tools like Poseblocks (a set of Scratch extensions that involve movement and dancing to make pose classifiers), Machine Learning for Kids (where children can use transfer learning to train their own classifiers~\cite{jordan2021poseblocks}), and Google’s Teachable Machines (an interactive tool that uses transfer learning to allow students to train a classification model using their own datasets~\cite{carney2020teachable}). Researchers have also engaged with teachers to bring AI learning to classrooms by training teachers to deploying these tools and curricula ~\cite{williams2022ai+, lee2022preparing} or co-designing with teachers to tailor these curricula for their students~\cite{lin2021engaging}.  

\subsection{Ethical Considerations of AI Systems}
Among motivators of promoting AI learning for K-12 children, a commonly stated one has been to make children aware of how AI impacts our society and their life. Touretzky et al. state that “students should understand that the ethical construction of AI systems that make decisions affecting people’s lives requires attention to the issues of transparency and fairness”~\cite{touretzky2019envisioning}. In this section we present some of the salient ethical issues that are currently being discussed.  

The Stanford Encyclopedia of Philosophy provides a summary of a breadth of work from the burgeoning field of AI ethics surfacing prominent debates about AI systems with notable potential for harm \cite{sep-ethics-ai}. These include: Privacy \& Surveillance, Manipulation of Behavior, Opacity of AI Systems, Bias in Decision Systems, Human-Robot Interaction, Automation and Employment, Autonomous Systems, Machine Ethics, Artificial Moral Agents, and Singularity. These provide a wide enough basis accessible for key conversations around how a variety of AI systems impact people and society, and what are different ways to respond to the same. 

A large survey of prior work around the responsible development of AI systems highlighted eight themes: privacy, accountability, safety and security, transparency and accountability, fairness and non-discrimination, human control of technology, professional responsibility, and promotion of human values~\cite{fjeld2020principled}. These themes have emerged from analyzing principles that are highlighted as critical tenets for the field, across works spanning different contexts, countries, and goals. These themes and principles are productive points for analyzing existing systems, anticipating harms of systems under development, crafting practices to inculcate across the field that can help mitigate potential harms, and also design and propose policy that can help protect users and consumers from harms relating to these topics. For instance, the summary document describes the theme of privacy as being described through the following principles: consent, control over the use of data, ability to restrict processing, right to rectification, right to erasure, privacy by design, and data protection laws. 

\subsection{AI Ethics Learning for K-12}
Ethical concerns pertaining to everyday AI systems also influence children's lives. Middle and high schools students interact with social media tools such as TikTok that use targeted advertisements, recommender systems for their social feeds, and face filters, or photo apps that use classifiers, face recognition algorithms and beauty filters. These students will soon apply for schools that could use AI-powered application sorters for admissions, or jobs that use automated resume sorters. As discussed in the previous section, these algorithms have social and ethical implications such as algorithmic bias and can influence childrens’ lives. In other words, AI is all around them and learning about its effect on their lives is critical. 

There are several existing AI literacy efforts for middle and high schoolers that focus on the ethics of AI systems. Researchers at MIT worked with teachers to designed an AI+ethics curriculum for middle schoolers where they take an integrated ethics approach and AI systems are taught through a socio-technical lens~\cite{williams2022ai+}. Technical concepts such as supervised learning or generative algorithms are accompanied by their ethical implications in lessons. For instance, students use learn about classification algorithms followed by how biased datasets can lead to discriminatory systems. In the DAILy curriculum, researchers integrated ethics and career futures with technical learning of ML concepts where students reflected on positive and negative impacts of technical systems on society~\cite{lee2021developing, zhang2022integrating}. The AI+Ethics curriculum for middle school students developed by MIT uses unplugged activities where students reflect on stakeholders of AI-powered tools and their needs and redesign these tools using principles of responsible design~\cite{ali2019constructionism}. The Creativity and AI curriculum proposes a generative models learning trajectory (LT) for young learners with a focus on Generative Adversarial Networks, creation and application of machine-generated media, and its ethical implications, and they pose these technical concepts as complex socio-technical concepts in this LT~\cite{ali2021exploring}. Some learning activities also focus on the ethical implications of just one kind of AI algorithm. For instance, \cite{ali2021children} discuss Deepfakes, or the generation of fake faces, and how it can be used to spread misinformation. 

Previous curricula and lessons showed that students had expressed surprise about the ethical implications of AI systems and gained new knowledge about how they affect their lives. Existing work, however, only exists in the form of learning activities in AI or ML curricula that can only be taught in classrooms or through online self-learning tools and make limited facilitation for discussion. Surfacing critical reflection is a key avenue of learning across contexts and disciplines~\cite{mezirow1990critical}, especially in learning environments aimed at developing skills to critique social conditions in ways that empower learners~\cite{friere1970pedagogy}. We conceptualized a fun way of discussing AI+Ethics concepts through a game. We learned from the benefits of game-based learning and designed a game that affords critical reflection on the ethical implications of AI through competitive play and discussion. 

\subsection{Game-based Learning}
Educational games have been used in a variety of contexts for their ability to enable greater excitement, engagement and deeper learning in numerous concepts~\cite{squire2014video}. Games often provide players with experiences where they make decisions within a simulated system, and strategize about the effects of different decisions within this system. This experimentation allows for a unique form of communication called procedural rhetoric~\cite{bogost2008rhetoric} – where the “consumer” of games is explicitly actively involved in making decisions which develops their understanding, and not just interpreting a statement given by others, as is the case with most traditional media and instruction. This experience often allows for richer experimentation and experiencing different kinds of failure in low-stakes environments – supporting a personally developed understanding of the system that is both fun, and a more involved engagement with different concepts that are typically provided by textbooks and other traditional educational media~\cite{gee2007good}. 

Another key avenue of learning enabled through games is in the form of diverse social engagement and consumption. Over 60\% of players participate in offline or online social communities related to their gameplay \cite{picton2020video}. A key positive impact of this is noticed in how reading in the context of games improves literacy skills among youth, since it takes place in a context that they are personally and socially invested in~\cite{steinkuehler2010reading}. Antle's description of emergent dialogue  ~\cite{antle2014emergent} provides another example of learning in games through social decision making. In a digital-analog hybrid tabletop game they designed, players learned through system-embedded decisions about ecological phenomena, as well as through designed in-game events' moments requiring critical reflection, talking to each other, bridging conflict and making collective decisions.

To make these conversations accessible across classrooms without needing additional technology, we also decided to design this game as an analog card game. This is additionally in response to the increasing recognition for the value of more unplugged activities, especially in computing classrooms. Such activities create space to think about computing concepts in ways that are not over reliant on specific technologies. Unplugged activities tend to be more accessible in classrooms – in terms of setup time, as well as cost and resources needed for implementation~\cite{bell2009computer}. Primarily, unplugged activities can enable richer interpersonal conversation and dialogue in classrooms which in turn can provide for rich collaborative and social learning opportunities. Even without intentionally designing for computing concepts, complex board and card games often elicit computational thinking principles in their rulesets and gameplay~\cite{berland2011collaborative}. 

Synthesizing these key aspects of fostering critical conversations through engaging and accessible classroom experiences, we chose to design an easy to print, setup and play analog card game as a powerful medium for interpersonal conversation and reflection around the ethical impacts of AI systems. 

\section{Design}

\subsection{Overview}
This game is themed around running businesses centered around different AI systems and technologies, challenging others’ businesses for potential harms they might be causing, and defending one’s own businesses by using features which can mitigate or remove these potential harms. Players play as business owners that set up new businesses - an attempt to replicate the real world creators of technology who are reflecting on the ethical implications of the AI-powered businesses they are creating. We focussed on our intended audience of middle and high schoolers while thinking about AI technologies and businesses that would be familiar to them, and have implications on their life for instance, recommender systems for social media, and face filters for camera apps. In addition to our prior expertise and media awareness, we leaned on the Stanford Encyclopedia of Philosophy’s debate highlights~\cite{sep-ethics-ai} to come up with descriptions of harms that are both easy to grasp for youth, and map to different subsets of the businesses we came up with (for instance, manipulating people’s buying behaviors coming directly from the debate around manipulation of behaviors which can happen in a wide plethora of ways from AI systems). Lastly, we came up with features that can respond to mitigate different harms we picked for the game, which also significantly overlapped with the key themes highlighted by Fjeld et al. as important considerations for the creation and assessment of AI systems (For instance, “Making the underlying AI technology and data usage transparent” as a feature in our game clearly corresponds to the theme of Transparency and Accountability). 

\textbf{AI Concepts Addressed \& Expected Learning Outcomes:}
Students will be made aware of real-world examples of harms caused by commonly available and used AI systems and examples of design and development practices AI creators can implement to mitigate negative impacts of their creations. Students will role play as AI creators and founders and practice thinking critically about the societal implications of the businesses they want to set up and how they can responsibly design their business.

\subsection{Game Materials}
The game consisted of three kinds of cards: Business Cards, Harm Cards and Feature Cards, listed below along with the game rule connections describing which harms can be caused by different businesses, and which features can address different harms. 

\begin{figure}[t]
\centering
\includegraphics[width=1\columnwidth]{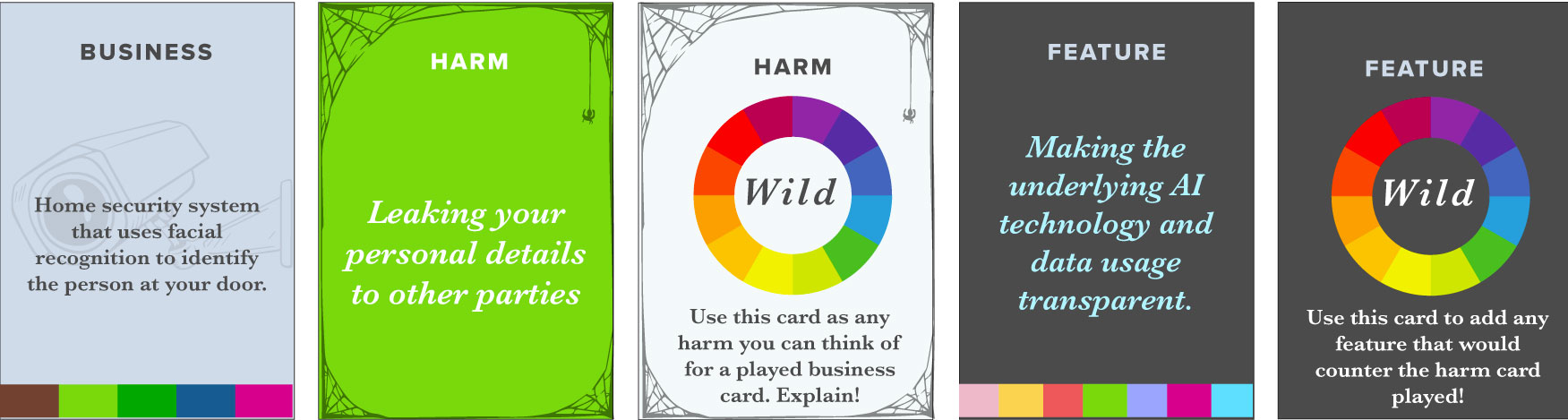} 
\caption{A Business Card, a Harm Card, a Wild Harm Card, a Feature Card and a Wild Feature Card}
\label{figcards}
\end{figure}

\textbf{Business Cards:}
\begin{enumerate}
    \item Home security system that uses facial recognition to identify the person at your door. \textit{Harms: 5, 8, 10, 11, 13}
    \item Crime prediction tool that can predict future crimes one week in advance with about 90\% accuracy. \textit{Harms: 7, 8, 10, 11}
    \item Personalized advertisement technology on websites people browse. \textit{Harms: 5, 6}
    \item Hiring algorithms that automate hiring in big companies to reduce the time taken to go through thousands of resumes. \textit{Harms: 3, 7, 8, 12}
    \item College admissions automator that decides who should be admitted based on different aspects in their application. \textit{Harms: 7, 8, 12}
    \item Self-driving cars. \textit{Harms: 7, 8, 10}
    \item Conversational agents. \textit{Harms: 2, 3, 4, 5, 6}
    \item Language translation algorithm. \textit{Harms: 2, 7, 8}
    \item Medical imaging to detect skin cancer from face images. \textit{Harms: 7, 8, 13}
    \item Recommender system for social media apps that personalizes your homepage’s feed. \textit{Harms: 1, 2, 3, 4, 5, 6}
    \item Generative AI Art magazine. \textit{Harms: 2, 7, 8, 11}
    \item Face filters people can use to apply different styles to their face. 	\textit{Harms: 1, 8}
    \item Social interactive robot. \textit{Harms: 1, 5, 6, 13}
    \item Personalizing search engine results to give you results specific to your past searches. \textit{Harms: 1, 2, 3}
\end{enumerate}

\textbf{Harm Cards:}
\begin{enumerate}
    \item Increased mental health challenges like depression, body dysmorphia, eating disorders.
    \item Spreading misinformation.
    \item Forming filter bubbles that isolate unique opinions from one another.
    \item Encouraging hateful behavior and hate groups.
    \item Leaking your personal details to other parties. 
    \item Manipulating people’s buying behaviors. 
    \item Taking over existing human jobs. 
    \item Algorithmic bias discriminating people based on their race, gender, ethnicity, or occupation.
    \item Misdiagnosing a patient’s illness. 
    \item Over-Policing neighborhoods.
    \item Leading to wrongful arrests of people.
    \item Marginalizing populations already under-represented in the workforce.
    \item Overly placing trust in imperfect technology.
\end{enumerate}

\textbf{Feature Cards:}
\begin{enumerate}
    \item Making the underlying AI technology and data usage transparent and explainable to users. \textit{Harms: 1, 2, 3, 5, 6, 9, 13}
    \item End to end encryption of data collected. \textit{Harms: 5}
    \item Collecting a balanced, diverse and large dataset to train the AI technology to reduce algorithmic bias. \textit{Harms: 3, 8, 11}
    \item Enabling people to control the degree of automation in their tools. \textit{Harms: 3, 6, 9, 10, 13}
    \item Employing a diverse team to develop this technology to gain diverse perspectives and address diverse needs. \textit{Harms: 1, 4, 7, 12}
    \item Including all affected populations of the technology in the design of the system. \textit{Harms: 1, 7, 12}
    \item Decision making by AI technologies to be examined by humans in the loop. \textit{Harms: 2, 7, 8, 9, 13}
\end{enumerate}

Additionally, we also created a Wild Harm Card and a Wild Feature Card (Figure 1), which were intended to provide creative space to players to come up with their own harm that a business might be causing, or their own feature that can mitigate a harm challenge played against them. All game materials are provided for anyone to download and print~\footnote{https://bit.ly/3BxW3M1}.

\subsection{Gameplay Rules}
This game is designed to be played by up to 7 players at a time, alongside educator facilitation depending on learners’ prior experiences with AI and technological literacy and comfort in engaging in discussion.
AI Audit consists of 3 kinds of cards: Business (14), Harm (13 x 3), Feature (8 x 2). The aim of the game is to choose and run a technology business which is able to implement appropriate features to mitigate causing harms. The player with the last surviving business wins. 

\subsubsection{SETUP:}
The game is set up by distributing all the Business cards across all players evenly (returning any extras to the box), and arranging the Harm and Feature cards face down in two separate decks. Players decide who goes first and play follows clockwise order. In the first round, each player draws 2 Harm and 3 Feature cards randomly (keeping all cards in their hand secret from others), and plays 1 Business card from their hand face down in front of them – the action of setting it up.

\subsubsection{TURN:}

\begin{figure}[t]
\centering
\includegraphics[width=0.7\columnwidth]{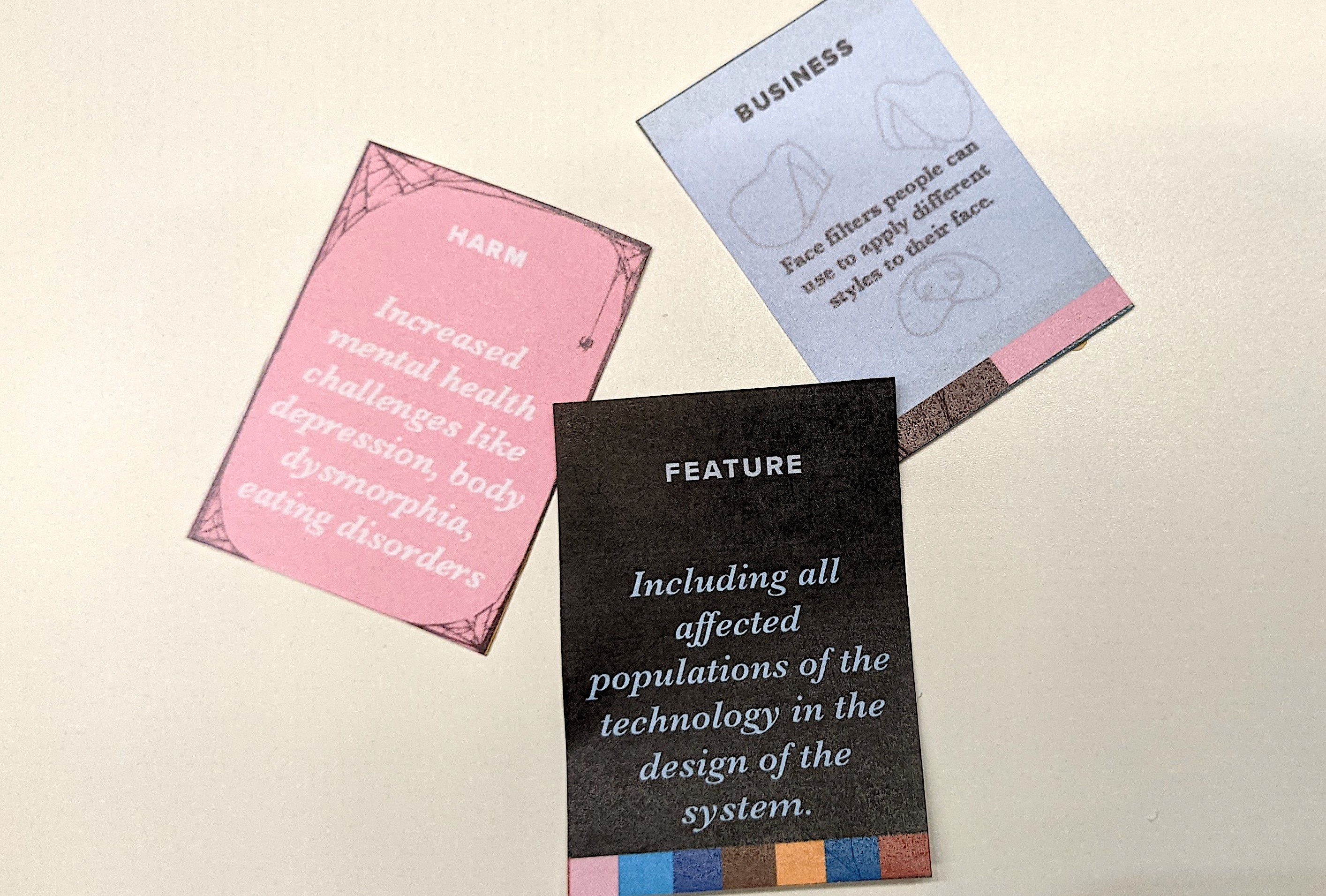} 
\caption{An example turn from the gameplay}
\label{figturn}
\end{figure}

After this setup, in each turn, a player can either set up a Business or play a Harm card against another player’s Business. Each Business card has a set of colors indicating which harms it can cause. A player can only play a Harm card against a business if their Harm card’s color is in the list of colors corresponding to the business. 

This challenge should be accompanied with a more detailed description of what this harm looks like. This description can come from players’ own understanding and intuition, the accompanying materials that players are provided as a part of the game, or from the supporting teacher to share explanations of what these harms look like for different businesses. While not mandatory for gameplay, we expect this conversational space to be the key venue for players to engage in conversation and deepen their understanding about such businesses and technologies, and the harms they can cause. 

Each challenged player gets a chance to respond to the challenging Harm with a Feature card. Similar to Business cards, Feature cards also have a set of colors indicating which Harms they can respond to. Players should only play Feature cards if their colors match the color of the Harm card played against them. If the challenged player has a matching Feature card and plays it, the Harm challenge fails – both the Harm and Feature card go to the bottom of their corresponding decks, and both players get to draw a replacement Harm and Feature card (Figure 2).

This is another juncture where players are encouraged to provide a more detailed description of how they think the Feature actually obviates the Harm that their business might have been causing. As in the case above, this description can come from players’ own understanding, accompanying educational materials, or accompanying educator support. Wild Feature cards can be played against any harm, but should be accompanied by a narrative on how it obviates the challenging Harm. Being able to convince a majority of the players that their imagined Feature is appropriate for the Harm challenge enables them to succeed against the Harm challenge.

    If players do not have an appropriate Feature to play against a Harm, the Harm challenge succeeds and they lose the challenged Business which goes to the discard pile. If a player has no running Business in their turn, they have to set up a Business. They can also choose to set up a second or third Business in their turn if they want to. 

\subsection{Educational Materials:}
The game is accompanied with an AI audit guide: including the gameplay rules outlined above, and real-life instances to describe the relations between different AI businesses and all their in-game corresponding harms. We expect these to be learning resources and conversation starters for students. The following is an excerpt from the AI audit guide: 

\begin{quote}
\textbf{Hiring algorithms that automates hiring in big companies to reduce the time taken to go through thousands of resumes. 
}

\textit{Algorithmic bias discriminating people based on their race, gender, ethnicity, or occupation: }
Resume sorters often make use of historical data with demographic information to make decisions about new data. This historical data might often have algorithmic biases, or might prefer candidates based on their race, gender, economic status or even their name. A recent study found that hiring algorithms are more likely to select applicants with common white names like Emily or Greg, versus distinctively Black names like Jamal or Lakisha.

\textit{Taking over existing human jobs:}
Replacing a human recruiter with an automated hiring system may be time efficient, but what happens to the human recruiter's job? Is it now redundant? According to a recent survey, companies are increasingly adopting AI powered screening tools for the first round of resume sorting, dramatically altering human recruiters’ jobs. 
\end{quote}

When played in classroom settings, the AI Audit guide is also a useful resource for teachers to facilitate discussions and provide concrete examples from real life to support the Harm and Feature cards. 

\section{Initial Testing}

\subsection{Playtesting}
We conducted an initial playtesting of the card game with five adults (four graduate students and one professional; age range 21 to 30 years) accompanied by one observer. Two out of the five players had prior expertise with AI. While this player group did not reflect our game’s intended player age group, it gave us initial playtesting feedback to improvise the gameplay. The observer used the AI Audit guide booklet to read out the rules to the players. We observed for: (1) challenges that players had in understanding the gameplay, (2) whether gameplay went as intended, (3) how engaging the game was, and (4) players’ reflections on AI ethics concepts after the game. players were informed that the game design is in progress and any feedback they have would be helpful for iterating the game. Post gameplay players were asked the following questions: 

\begin{itemize}
    \item Did you enjoy the game? Why or why not?
    \item Did you have any trouble understanding the game rules? 
    \item Did you learn anything new about the societal and ethical effects of AI? 
    \item What strategies did you employ while playing the game?
    \item What modifications would you suggest for the game? 
\end{itemize}

\subsection{Results}
Players had no trouble understanding the gameplay and found it straightforward. The only question asked before beginning the gameplay was “What happens if I don’t have an appropriate harm card and all business [cards] are played?” where we told them to pass the turn in that case. Players chose to play the game three times. On average, a game round lasted for 19 minutes and 40 seconds (R1: 14 min, R2: 34 min, R3: 11 min).

The gameplay sparked several discussions about ethical impacts of AI technologies. In the first round players referred to the AI Audit guide more than in future rounds. For instance, one player who had the “Generate AI Art Magazine” business was confused about how it can lead to “Wrongful arrests of people”. When they referred to the guide, they learned that generative AI can be used to create Deepfakes, or modified videos of people and create false evidence presentable in court. The player was surprised to learn about this possibility. 

One major limitation in the gameplay was that since the businesses in a game round are fixed, there came a point in round 2 where all players’ harm cards were rendered redundant since they did not match any of the business cards. At this point players decided to switch a Harm card in their hand for one from the Harm cards' deck to make the game progress. Players also noticed how it is very easy to keep countering Harms because the Feature cards are so powerful, so they decided to discard one Feature card each and move to two. Players also struggled with some terminology. For instance, one player asked what a “conversational agent" is. One player was not sure how face filters use AI. All players commented on how close some of the colors were, making them difficult to differentiate, like pink vs red. Players frequently expressed surprise and interest in the Features. For instance, when the Feature “End to end encryption of data collected” was used to counter the harm “Leaking your personal details to other parties” for the business “Personalizing search engine results”, one player said, “Oh that’s so interesting.” players were also excited about their strong features. For instance, one player was very proud that they will be “employing so many new and diverse people in their business”. While some players viewed the wild card as a replacement of any other Harm or Feature card in the game, some viewed it as a Harm or Feature they can conceptualize from even outside the given cards (the intended design). 

In the post game feedback, players reported that the game was very fun and competitive. players suggested making the colors more distinct. One player mentioned that the text on some cards may be too long or complex for children and to make it simpler. players mentioned how the examples in the guide were very helpful. One player said that they kept getting the same feature and wanted unique ones. When asked about challenges, one player mentioned that they were not sure how to use the Wild Card and it would be helpful to have an example, especially for younger players. One player said, “I like the Wild because it is powerful, but I don’t like it because it makes me think so much.” Given the competitive nature of the game, Feature Wild Card players also had to come up with a strong argument to defend their card. 

When asked about what they learned, one player said, “I didn't know that facial recognition can lead to wrongful arrest”. Another said, “I learned that Self-driving cars are not a good business and social media can lead to depression”. One player said, “Made me think about something I would normally not pay attention to. Like I knew it, like the targeted ads, but I don't think about it on a daily basis. But when I was putting a business that is gonna create advertisements for people, I would need to create an algorithm that will be diverse and will reach more people.” Among strategies that players employed were putting down the Businesses vulnerable to similar sets of Harms so one can act as back-up in response to used up Harm challenges; trying to play Businesses similar to other people to increase the chances of a particular Harm challenge being played against other players; and trying to play the least harmful businesses first.

\subsection{Design Changes}
Several changes emerged from this initial round of testing.
\begin{itemize}
    \item Simplifying the language of some Business cards and adding familiar words. For instance, “Conversational Agents” can be changed to “Voice Agents like Alexa”. Further, we aim to make it more explicit how these businesses involve AI and make the text length shorter. 
    \item Colors on the cards can be made more distinct, and can be accompanied by distinct shapes to make them easy to differentiate. As one player pointed out, this will also be friendly for players with color-blindness. 
    \item Adding an example of Wild Card usage in the AI Audit guide booklet. 
    \item Only providing two Harm and two Feature cards to make the game progress faster.
    \item Once all players have redundant Harm cards, they can exchange their Harm cards for new ones from the pile. 
\end{itemize}

\section{Discussion}
Our design context and designed game provides a very promising ground for developing similar games that foster rich critical conversations about different socio-technical systems relevant in youth lives. Noticing even our adult playtesters being surprised by different connections between technologies and harms, and also ways to respond to potential harms, surfaced a key value of how games enable just in time and contextualized learning ~\cite{gee2007good}. Without needing players to remember or have to be constantly imaginative in coming up with harms or solutions in the game, the different cards provide a rich scaffold for players to learn. 

At the same time, we see critical value in making space for open-ended conversation and storytelling that built on these scaffolds – players don’t need to come up with a “correct” answer which introduces space for failure in case of lack of knowledge, but more simply create narratives for connections that the game already provides. This personal narrativization connects to extensive work on the power of storytelling in enhancing learning ~\cite{alterio2003learning}, and is an instance of overtly surfacing constructivist learning practices – wherein people learn more effectively by having opportunity to create and solidify their own understandings of concepts by drawing on their own prior experiences and knowledge \cite{piaget1967logique}.
Additionally, the wild cards provide some key avenue for fuller creative thinking and storytelling by the players. Leaning on the structure of educational scaffolds ~\cite{melero2011review}, the wild cards act as a space with lesser imposed structure inviting players to engage in the practice of understanding the broader systems, anticipating impacts of AI systems, and improving these systems to respond to possible harm. We expect that creating this bridging experience within the game itself will be hard, it is a critical opening to connect to the broader learning goal of thinking critically about socio-technical systems, understanding their working and how to critique as well as improve them.

Our pilot game design work as well as playtesting frequently brought up the potential to integrate richer game mechanics around business management and economic decisions to balance ideas of longevity, investment, costs of different services. The possibility of engaging players in game and system management complexity presents a tension against keeping the game mechanics minimal to focus on only engaging around critical conversations. While we hope to try different game designs in future work, we believe prioritizing a narrower focus on the salient mechanic of critical discussions has helped make a game that is quite easy to learn, set up, and play. This is particularly suited for games intended for classrooms and similar time bounded contexts. 

Relatedly, being able to design the game with educator support in mind has helped our game design as well as creation of supporting material. Especially, choosing to make a multiplayer game enables us to place the educator as almost a co-player, which increases the flexibility of our design and the capacity with which different kinds of players can engage with the material at hand. While this design still needs more testing, we strongly recommend the design of learning games to increase space for co-participation from supporting teachers, which is best supported in multiplayer games, and more so in analog games and contexts which involve more direct interpersonal interaction.

Finally, while existing works in K-12 AI Ethics literacy enable children to critique existing AI technologies and identify their potential harms, this work allows them to put themselves in the shoes of the creators of these technologies and conceptualize their responsible redesign, while being supported by learning materials that scaffold their knowledge about AI-enabled societal harms. 

\section{Conclusion}
We designed a card game for middle and high school students to reflect on the ethical implications of everyday AI systems. The competitive game facilitates children’s learning about how AI-powered businesses can have negative impacts and how these businesses can mitigate those impacts. The game is designed to facilitate conversation between the players to critically reflect on and discuss their thoughts about how various technologies harm us. The cards are scaffolded in a way that aims to make children aware of how their everyday technology can potentially be harmful in ways that they do not anticipate. The goal is to help them be responsible consumers and creators of AI. In our initial playtesting with adults, we found that the game was engaging and fun, even for adults, but players also found some content challenging. We found that the accompanying education materials were helpful and that players gained new insights about societal implications that they were not aware of, or were reminded to think critically about the ones that they were aware of.  

\subsection{Future Work}
In future work, we aim to test the game with middle and high school students and teachers and observe the kind of discussions students have. We will observe how difficult or easy the game and its terminologies are for different age groups. We also aim to observe the game in a classroom setting and  players’ and teachers’ roles in engaging with the supplementary AI audit guide and facilitating gameplay.  Future versions of the game could involve different difficulty levels for different age groups. There could also be variations with more wild cards to allow for deeper critical thinking about the harms of AI technologies. To increase the replayability of the game, we could also release new business card decks that can be used in addition to the original game. This work could also inspire the use of reflective games for discussing other socio-technical systems outside of AI. Finally, we aim to utilize AI ethics assessments to analyze how the game benefits children’s gain of AI ethics knowledge. 

\bibliography{paper}

\end{document}